\documentclass{JHEP3}
\preprint{LU-TP 05-25,\\
  LUNFD6/(NFFL-7223)2005\\
  hep-ph/0505181}
\usepackage{cite}
\usepackage{epsfig}
\usepackage{color}
\let\normalcolor\relax
\usepackage{graphics}
\usepackage{axodraw}
\usepackage{inputenc}
\usepackage{xspace}
\inputencoding{latin1}
 \renewcommand\email[1]{{\scriptsize\tt\href{mailto:#1}{#1}}}

\newcommand{\pT}{\ensuremath{p_{\perp}}}

\newcommand{\pythia}{P\scalebox{0.8}{YTHIA}\xspace}
\newcommand{\charybdis}{C\scalebox{0.8}{HARYBDIS}\xspace}

\def\mrm#1{\mathrm{#1}}
\def\sub#1{\ensuremath{_{\mrm{#1}}}}
\def\sup#1{\ensuremath{^{\mrm{#1}}}}

\newcommand{\eqref}[1]{eq.~(\ref{#1})\xspace}

\def\rsch{\ensuremath{r\sub{Sch}}}
\def\Et{\ensuremath{E_\perp}}
\def\shat{\ensuremath{\hat{s}}}
\def\sighat{\ensuremath{\hat{\sigma}}}
\def\mpl{\ensuremath{M\sub{P}}}
\def\mpln#1{\ensuremath{M\sub{P #1}}}
\def\mmin{\ensuremath{M\sub{min}}}
\def\ie{i.e.\ }
\def\eg{e.g.\ }

\skip\footins = 1\bigskipamount plus 2pt minus 4pt                              

\title{\boldmath QCD-supression by Black Hole Production at the LHC}

\author{Leif Lönnblad and Malin Sjödahl\\
  Dept.~of Theoretical Physics,
  S\"olvegatan 14A, S-223 62  Lund, Sweden\\
  E-mail: \email{Leif.Lonnblad@thep.lu.se}
    and \email{Malin.Sjodahl@thep.lu.se}}
  
\author{Torsten Åkesson\\
  Dept.~of Physics,
  Lund University,
  Box 118, S-221 00  Lund, Sweden\\
  E-mail: \email{Torsten.Akesson@hep.lu.se}}

\abstract{Possible consequences of the production of small black holes
at the LHC for different scenarios with large extra dimensions are
investigated. The effects from black hole production on some standard jet
observables are examined, concentrating on the reduction of the QCD cross
section. It is found that black hole production of partons interacting on a
short enough distance indeed seem to generate a drastic drop in the QCD
cross section. However from an experimental point of view this will 
in most cases be camouflaged by energetic radiation from the black
holes.}

\begin{document}
 
%set sloppy attitude to line breaks

Since the theory of Large Extra Dimensions by Arkani-Hamed Dvali and
Dimopoulos (ADD) first saw light in 1998 \cite{Arkani-Hamed:1998rs}
there has been much discussion in the literature about the prospect
for observing these large extra dimensions in colliders (see \eg
\cite{Antoniadis:1998ig,Emparan:2000rs, Dimopoulos:2001hw,Giddings:2001bu,Kanti:2004nr}). In
particular the signal of small black holes which can be produced if
the Planck scale, \mpl, is of order TeV has been discussed.  However,
there is another characteristic of black hole events; the signal of no
signal\cite{Banks:1999gd} .  
%That such possibilities could exist has also previously been
%suggested by \cite{Banks:1999gd,Giddings:2001bu,Hofmann:2002xd}. 
If we \eg consider the dijet cross section as a function of
invariant mass, $\sqrt{\hat{s}}$, then for some large invariant mass,
the partons which otherwise would have undergone a QCD scattering may
instead be trapped in a black hole which would lead to a falling dijet
cross section as a function of invariant mass.

This argument is clearly oversimplified. Requiring a large $\hat{s}$
ensures a small extension of the interaction in beam direction.
However, to form a black hole the energy has to be well confined also
in transverse direction.  The absolute range of a QCD interaction is
set by the virtuality of the propagator, $Q^2$.  This sets the
timescale of the fluctuation and we therefore expect that interactions
with a large enough virtuality takes place over a small enough space
such that it is really possible to form a black hole, at least if the
mass is high enough to avoid uncertainties due to quantum gravity.
Hence, the dijet cross section would only fall off for high enough
\Et\ of the jets. Of course, some of the produced black holes may
themselves decay into hard partons, filling up the jet spectrum at
high masses and \Et, but we expect that spectra to look quite
different from standard QCD.

% \todo{Here we should include something about the decay of black holes}

For energies much larger than the Planck scale, corresponding to a
Compton wavelength much smaller than the Schwarzschild radius of a
black hole with that energy, we expect classical gravity to give the
correct result, and a produced black hole would evaporate by emitting
particles with the thermal spectra from Hawking radiation. The problem
is that in colliders we would first probe the Planck scale region
where quantum gravitational effects should come into play.

Approaching the Planck scale from below, where gravity starts to
become important, but still is not strong enough to trap partons in
black holes, we would expect gravitational scattering, to leading
order simply $2\to2$ partonic
scatterings\cite{Han:1998sg,Atwood:1999qd,Doncheski:2001se}. As the
energy increases, the gravitational scattering would transform into
small black hole events. On the other hand approaching from above
using the approximation of Hawking radiation we would expect the black
holes to decay to \textit{a few} particles.

In this transition region we are sensitive to unknown quantum gravity
effects, but from general continuity arguments we would expect the black
holes to decay into a small number of particles, preferably two. Then
as the energy increases we expect the black holes to emit more and more
particles looking more and classical and, in the end for $M \gg\mpl$,
we would expect the thermal spectra from Hawking radiation.

Hence, in searching for a drop in the cross section, a more
complicated jet spectra will appear with increasing $E_T$ scales.
First there will be an increase of the cross section due to
gravitational interactions. Then, as black holes begin to form, they
are expected to decay to relatively few particles which will sometimes
be identified as jets, resulting in an increase of the cross section.
At high enough jet scales, however, we expect the jet cross section to
completely die out, since the QCD component of the hard scattering
will disappear and the black hole decay products will be softer as the
black hole masses increase.

From the point of view of searching for large extra dimensions it is,
of course, the increased cross section which will be easiest
to see, already at scales much below where black holes are formed. But
an increase in the jet cross section could be attributed to many
different kinds of \emph{new physics}, while the disappearance of
Standard Model cross sections is not normally expected from new
physics processes.

So, while the above mentioned effects (gravitational scattering and
black hole decay products) are well worth looking for, we will devote
this paper to the black hole 'eating' of the QCD cross section, and we
will do this by using phenomenological event generator models. We
intend to get back to the other issues in future publications.

I this paper we start with a brief review of the ADD and the, somewhat
different, Randall--Sundrum (RS) \cite{Randall:1999ee} extra-dimension
scenarios in section \ref{sec:models-large-extra}. The model for
generating and decaying black holes is described in section
\ref{sec:Procedure}, while the many theoretical uncertainties are
discussed in section \ref{sec:Uncertainties}.  In section
\ref{sec:results} we present and comment on our results and in section
\ref{sec:conclusions} we make the concluding remarks and suggest
further studies.

%Hence we will here treat the additional gravitational scattering as something 
%we like to get rid of. The problem is, as mentioned above, that we expect 
%these events to be rather similar to other dijet events as the partons still 
%have color so there will still be a color field stretched by the 
%gravitationally scattered partons. As the partons move apart we expect the 
%color fields to stretch and break as in normal hadronization in a QCD event. 
%\footnote{We do expect a statistical difference from QCD scattering since the 
%color flow will be modified as the graviton does not carry color.} The way to 
%eliminate these extra scatterings will instead be by choosing to look only 
%at processes with large $p_T$ for a given $\hat{s}$. For a large enough $p_T$ 
%implying a small enough interaction region we will have a black hole.  

\section{Models of large extra dimensions}
\label{sec:models-large-extra}

\subsection{Basics of ADD}
\label{sec:ADD}

The aim of the Arkani-Hamed Dvali and Dimopoulos (ADD) model
\cite{Arkani-Hamed:1998rs} is to solve the hierarchy problem of the
differences in scale between the electro-weak scale, 100 -- 1000 GeV,
and the scale where gravitation becomes important, $10^{19}$ GeV.
This is done by introducing large extra dimensions with some
compactification radius, $R$.  Although the compactification radius is
typically taken to be the same in all extra dimensions it could in
principle vary. For distances much smaller than the size of the extra
dimensions (here assumed to be the same) we will then have a Newton's
law of the form \cite{Arkani-Hamed:1998rs}
\begin{eqnarray}
  \label{eq:Vsmallr}
  V(r) \sim \frac{M}{\mpl^{n+2}}\frac{1}{r^{n+1}},
\end{eqnarray}
where $n$ is the number of extra dimensions and \mpl\ the fundamental $n+4$
dimensional Planck mass. On the hand, for $r \gg R$
gravity has expanded to the full volume of the extra dimensions and we
get \cite{Arkani-Hamed:1998rs}
\begin{eqnarray}
  \label{eq:Vlarger}
  V(r) \sim \frac{M}{\mpl^{n+2}}\frac{1}{R^{n}}\frac{1}{r}
\end{eqnarray}
But this must equal to Newton's law in 3+1 dimensions so we conclude
that the observed four dimensional Planck mass \mpln{4}\ is (up to small volume
factors)
\begin{eqnarray}
  \label{eq:Mp4}
  \mpln{4}^{2} \sim \mpl^{n+2}R^n
\end{eqnarray}

This explains how we could have a fundamental Planck scale almost at
the electroweak scale but an observed Planck scale at $10^{19}$ GeV.
Requiring that the fundamental Planck scale is the same as the
electroweak scale, $\sim$ TeV, gives a compactification radius of the
solar system for one extra dimension.  This is trivially excluded from
observations.  Two extra dimensions gives $R \sim$ mm, which is
indirectly excluded from various cosmological and astrophysical
constraints.  The same holds for $R \sim$ nm, corresponding to three
extra dimensions.  However for four or more extra dimensions it is
still possible to have a fundamental Planck scale at the order of a
TeV (A more complete listing is given in \eg \cite{Kanti:2004nr}).

Since the black holes considered here are well within the range $r \ll
R$, the Schwarzschild radius can be calculated analogous to the 3+1
dimensional case. The result is \cite{Myers:1986un}
\begin{eqnarray}
  \label{eq:rSch}
  \rsch = \frac{1}{\sqrt{\pi} \mpl}
  \left[ \frac{M\sub{BH}}{\mpl} \frac{8 \Gamma(\frac{n+3}{2})}{n+2} \right]^{\frac{1}{n+1}}
\end{eqnarray}
where $\Gamma$ is the Euler gamma function.

The temperature is given by \cite{Myers:1986un}
\begin{eqnarray}
  \label{eq:T}
  T = \frac{n+1}{4 \pi \rsch}
\end{eqnarray}
This means that more massive holes are colder. The mass dependence is
however much weaker than in 4 dimensions since the radius changes less
with mass. The dependence on the number of extra dimensions is dominated 
by the factor $n+1$ in the numerator.
Hence black holes in many dimensions are hotter.

\subsection{Basics of RS}
\label{sec:RS}
A somewhat different model was introduced one year after the ADD-model
by Randall and Sundrum \cite{Randall:1999ee}.  In the Randall--Sundrum
model, RS, which has only one extra dimension, the metric doesn't
factorize. Instead the 4 dimensional metric is multiplied by a
``warp'' factor which depends on the coordinate, $y\in [-\pi,\pi]$, in
the fifth dimension.

\begin{eqnarray}
  \label{eq:metric}
  ds^2 = e^{-2|y|/l}\eta_{ij}dx^i dx^j+dy^2
\end{eqnarray}
In the above expression $\eta_{ij}$ is our normal (flat) four
dimensional metric with indices running from 0 to 3 and $l$ is the
radius in the anti-de Sitter space.  By taking the warp factor to be
an exponent, and by assuming the gauge fields (and hence us) to live
in the most shrinked slice, $y=\pi$, of the five dimensional world, a
large hierarchy can be accomplished by letting gravity propagate in
the full space.

We note that it has been suggested (see eg.\ \cite{Karasik:2003tx})
that black holes in the RS model can only be formed on the Planck
brane, which effectively means that there would be no black holes
produced in a collider. However, it was also suggested
\cite{Karasik:2004wk} that the naked singularity obtained instead,
could be covered on the Standard-Model brane. Such a singularity could
then be produced in a collider and would evaporate as gravitational
radiation into the bulk (where it is naked) at the Planck time scale.
As this radiation is not detected it would appear as missing energy.

%% If, as discussed in \cite{Karasik:2003tx,Karasik:2004wk}, no black
%% holes, but instead a naked singularity is produced in the
%% Randall--Sundrum scenario, the naked singularity would evaporate as
%% gravitational radiation into the bulk (where it is naked) at the
%% Planck time scale. As this radiation is not detected it would appear
%% as missing energy.
%\comment{Does the naked singularity have the cross section
% $\pi r^2$, and does it eat QCD cross section? }

On the other hand, if a black hole is formed, but is stable on
collider time scales as argued in \cite{Casadio:2001wh}, the hole
could (if gauge charged) be detected. It is, however, unlikely that
this would affect the signals studied here, and we have chosen to
ignore this effect.
% This may in principle give some 
% contribution to the signal studied here but we have ignored this effect.
% Most often, however, the black
% hole, taking over the momentum of the ingoing partons, would have a large 
% longitudinal but a small transverse momentum and hence most often escape 
% detection. It would therfore not contribute much to the signal studied here.
% \comment{More understandable?}.

If a RS black hole is sufficiently small, such that it is not affected
by the bulk curvature, the radius is still given by \eqref{eq:rSch}
\cite{Anchordoqui:2002fc}.  Hence, in either of the above mentioned
cases for black holes/naked singularities in the Randall--Sundrum
scenario, there will be no Hawking radiation which populates the spectra, but 
there will be a drop in total cross section for standard model interactions. 
Furthermore the cross section would be the same as
in the ADD scenario with one extra dimension for small enough black holes 
\cite{Anchordoqui:2002fc}.

For the purpose of this paper, we ignore all the uncertainties of
whether or not black holes are possible in the RS model. In fact, we
do not use any details of the RS model, but use it only as an example
of a model which could produce black holes/naked singularities at
large cross sections at a collider, but where the decay products of
the hole would not be detectable. For our investigations this is then
a best-case scenario.

\section{Black hole production and decay}
\label{sec:Procedure}

To study the effect on QCD-jets on LHC we have used
\pythia\cite{Sjostrand:2000wi} to generate the QCD events. To lowest
order we here have two partons with energy fractions, $x_1$ and $x_2$,
and an invariant mass, $\shat=x_1 x_2 s$, which scatter and give rise
to two outgoing partons with transverse momenta,
$\pT\approx\sqrt{Q^2}$. For this process to be trapped in a black
hole, we require the whole process to be located within the
Schwarzschild radius, $\rsch(\shat)$. Looking at the momenta of the
incoming partons in their combined rest frame we must require that their
wavelength, $\lambda_l\propto 2/\sqrt{\shat}$, is less than
\rsch. The corresponding requirement in the transverse direction gives
the requirement: $\lambda_\perp\propto 1/\pT<\rsch$.

Of course, it is questionable if these requirements are enough, maybe
the wavelengths should be \emph{much} smaller than \rsch. In any case
it is reasonable to introduce a parameter, and we will require
$\lambda<\rsch/P$, where we use $P=1$ as a standard value. From the
longitudinal requirement we then get a minimum mass of a black hole
from
\begin{eqnarray}
  \label{eq:Mmin1}
  \mmin=2P/\rsch(\mmin).
\end{eqnarray}
We then get a cutoff in the standard QCD cross section, given by the
following step-functions (which in general could be replaced by more
smooth suppression functions):
\begin{eqnarray}
  \label{eq:qcdcut}
  \frac{d\sigma\sub{QCD}(Q^2)}{dQ^2d\shat}=&\int& dx_1 dx_2
  \sum_{i,j}f_i(x_1,Q^2)f_j(x_2,Q^2)
  \frac{\sighat_{ij}\sup{QCD}(\hat{s},Q^2)}{dQ^2}\times\nonumber\\
  & &\delta(\hat{s}-x_1 x_2 s)
  \left[1-\Theta(\shat - \mmin^{2})
    \Theta(Q^2 - \frac{P^2}{\rsch^2(\shat)})\right].
\end{eqnarray}
where the sum runs over all parton types.

Instead of the QCD process we will get black holes with the cross section
\begin{eqnarray}
  \label{eq:sigma}
  \frac{\sigma\sup{BH}(\shat)}{d\shat} = \int dx_1 dx_2 \sum_{i,j}f_i(x_1,Q^2)
   f_j(x_2,Q^2) \sighat\sup{BH}(\hat{s})
   \delta(\hat{s}-x_1 x_2 s)\Theta(\shat - \mmin^{2}),
\end{eqnarray}
where $Q=P/\rsch(\shat)$. The partonic cross section is simply given
by $\hat{\sigma}=\pi\rsch^2$, but also here one could imagine a smooth 
transition and a factor in front (see section \ref{sec:Formation} below).

\subsection{\protect\charybdis}
\label{sec:charybdis}

We here describe the most important properties of the black hole event
generator \charybdis, which we have used to generate black holes and
their decays. The complete references are \cite{Harris:2003db,
  Harris:2003eg}.

In \charybdis two partons create a black hole according to
(\ref{eq:sigma})\footnote{By default the scale $Q^2$ is taken to be
  $\shat$ in \charybdis, but we have used the option with
  $Q^2=1/\rsch^2$ \cite{Giddings:2001bu}, since this gives a more
  continuous transition from \eqref{eq:qcdcut}.}, which then decays by
Hawking radiation.  The procedure of the decay is such that momentum,
charge, color and baryon numbers are conserved. The decay products are
chosen democratically among the standard model particles according to
their degrees of freedom (color and spin).  The energy of an emitted
particle is taken from a Planck spectra modified with the gray-body
factor of the particle. The temperature is given by \eqref{eq:T} where
the black hole mass by default is taken to be the remaining mass of
the hole, but this may be modified such that the initial temperature
is kept throughout the evaporation. The motivation for keeping the
initial temperature could be that is is anyhow dubious to treat the
black hole as a thermalized object considering the speed of the decay
\cite{Dimopoulos:2001hw}.

One problem is that even if a black hole would be created well
above the Planck scale, \ie well within the classical regime, it
would evaporate and eventually reach the Planck scale. And, although
we do not know what physics looks like at the Planck scale, there must
be a way of terminating the decay.

The \charybdis method for this is to have a free parameter, set by the
user, which gives the number of particles the hole decays to in the
end, where the end can be defined either as when the black hole reaches
the Planck mass, or - if it happens before - when one of the emitted
particles happens to get a momentum so large that energy and momentum
conservation would be violated for a two body decay.

\section{Uncertainties}
\label{sec:Uncertainties}

First of all it should be pointed out that the major lack in our
description is the absence of a theory for quantum gravity. As we do
not know what physics we will encounter at the Planck scale it is
impossible to make an error estimate in the traditional sense. However
we may still discuss which uncertainties will effect our results.
Roughly these can be classified in three groups. The first concerns
non black hole gravitational events, the second the production of
black holes, and the third deals with the decay of the black holes.

Apart from the uncertainties associated with quantum gravity there is of
course also an uncertainty in the QCD jet cross section. A recent evaluation
of the effect from the uncertainties in the parton distribution functions,
and in the NLO QCD theory, is published in \cite{Stump:2003yu}. 
In this paper a comparison is made between the calculated
cross section and the Tevatron Run 1b data, and the uncertainty in the jet
cross section for the LHC is estimated to be up to a factor 2.5 at a jet-ET
of 5 TeV. This effect is important for many QCD and other Standard Model
studies, however, it is an effect that can be neglected for the main 
conclusions in this paper as can be seen in e.g. figure 3.

\subsection{Non black hole gravitational events}
\label{sec:Grav}
As mentioned in the introduction it is highly unlikely that black
holes is the major gravitational effect. In the ADD scenario the
amplitude for perturbative scattering is calculated in
\cite{Han:1998sg} and the dijet cross section is spelled out (using
the Planck scale as cut of for the Kaluza Klein modes) in \eg 
\cite{Atwood:1999qd}. While the effect on jet production from
gravitational scattering is large, it has the disadvantage of depending
on a free parameter for the cutoff. Furthermore, as it is a
perturbative expansion it cannot be applied close in the Planck
region.  This means that when searching for the signal of a
disappearing QCD cross section, we have to do that in a spectra which
is already modified by gravitational scattering in a basically unknown
way. In this paper we will not further investigate this issue, and
these processes are not included in the simulation.
Although they should have a large effect on the cross section close to
the Planck scale, we will here mostly be concerned with the behavior
well beyond this scale.

% \todo{Gravitational effect in Randall Sundrum??}

\subsection{The formation of a black hole}
\label{sec:Formation}

There are several unresolved issues concerning the formation of mini
black holes.  Firstly there is an uncertainty already at the
classical level. The naive formula is $\sigma(M\sub{BH}) = \pi \rsch^2$.
Classical numerical simulations indicates that this value should be
multiplied by a factor $\sim 0.7-3$, increasing with the number of extra
dimensions \cite{Yoshino:2002tx,Yoshino:2005hi}. 
(For a discussion about the effects of quantum fluctuations 
based on wave packages, see \cite{Giddings:2004xy, Rychkov:2004sn}.) 
During the formation some energy may be lost as 
gravitational energy reducing the remaining mass for Hawking radiation 
\cite{Yoshino:2005hi,Cardoso:2005jq}.

Secondly the factorization of
the parton level cross section can be questioned since the process is
inherently non-perturbative \cite{Harris:2004xt}.
 
%and partly since we expect the extrapolation of the parton 
%distribution functions to be changed due to gravitational effects.

Another fundamental issue concerns the onset of black hole production.
One may argue that no black holes should be formed below \textit{roughly} 
the Planck scale since this is (approximately) the mass scale where the 
Compton wave length equals the black hole radius. 
This is the view point taken in \eqref{eq:Mmin1}. Combining \eqref{eq:Mmin1}
and \eqref{eq:rSch} we get the following relation for \mmin\

\begin{eqnarray}
  \label{eq:ratio2}
  \mmin={\mpl}\pi^{\frac{n+1}{2(n+2)}}
           \left[ \frac{n+2}{8 \Gamma(\frac{n+3}{2})} \right]^{\frac{1}{n+2}} 
  (2P)^{(n+1)/(n+2)}
\end{eqnarray}
We have used $P=1$ as a standard value in this paper. Numerically the value 
of \mmin\ is then approximately twice the Planck mass. 

As for $Q^2$, used in the parton distribution functions when generating 
the black holes, we like to use the same constant in the inverse relation 
between length and mass as in \eqref{eq:Mmin1} and this is what gave us 
$Q=P/\rsch$. While, in principle, the scale in the parton distribution 
functions and the minimal mass could be varied independently we have 
chosen not to.

%, as the choice for the PDF:s is much less important in the region of the onset of black hole production.  

One could also think of more sophisticated ways, where the nature of the 
force is taken into account, of relating the 'interaction distance' to 
momentum but that is beyond the scope of this article.

\subsection{The decay of a black hole}

It is common to divide the decay of a mini black hole into 3 different
phases. In the first the asymmetry of the black hole, due to energy
momentum and gauge charge distribution, is lost.  This is therefore
referred to as the ``balding phase'', and it cannot be incorporated
into \charybdis as the spectra of this phase is not known.
Furthermore, gravitational radiation can be emitted on the brane.  As
this radiation is not detected it will appear as a missing energy.
This radiation has also not been included in \charybdis, where black holes 
decay only into standard model particles.

The second phase is the Hawking evaporation phase. This is the phase
most accurately described by \charybdis. However there are still a few
discrepancies.  One is that \charybdis uses the spectra of a
non-rotating Schwarzschild black hole, while the differential cross
section would favor a Kerr hole with large angular momentum, see \eg
\cite{Giddings:2001bu}.
% \comment{This article explains it but I don't know if it was the first to do it}.  
A rotating black hole may emit a significant part of the total
radiation into the bulk
\cite{Page:1976ki,Frolov:2002xf,Frolov:2003en,Stojkovic:2004hp}.
There is not yet a consensus in the literature about how much is
emitted into the bulk, but it is not inconceivable that it is as much
as 50\%.  As large impact parameters are favored this may
significantly diminish the observed radiation.  Also, it may be
possible for the black hole to recoil off the brane, emitting further
Hawking radiation in the bulk \cite{Frolov:2002as,Frolov:2002gf}.

In principle the gauge charges of the black holes should also be taken
into account. In the case of no extra dimensions and larger black
holes ($M\sub{BH} \gg \mpl$) the suppression of charged particles due
to the electrostatic potential is argued to be a small effect of a few
per cent \cite{Page:1977um}. But since QCD is a significantly stronger
force, and since the black holes considered here are small, the effect
of QCD may be sizable. Again this is not properly included in
\charybdis.  There is, however, effectively a bias for events with few
charges since charge is conserved, implying that more events with large
charge emission are thrown away.

It can also be questioned if it is correct to treat these mini black
holes as thermalized considering their rapid decay
\cite{Dimopoulos:2001hw}\footnote{On the other hand the decay may be
slowed down as argued in \cite{Casadio:2001wh}.}. From the point of
view of this study a varying temperature is however the most
conservative choice since a hotter (thermalized) hole gives fewer
decay products and hence is more difficult to distinguish from the QCD
dijet background. We hence chose a time varying temperature. An
example of the difference between a varying and a non varying
temperature can be found in \cite{Harris:2004xt}.

Furthermore, it has been argued that the rapid decay may lead to the
black hole becoming surrounded by a \textit{chromosphere} of soft
partons which could suppress emission of hard
partons\cite{Anchordoqui:2002cp}. Such effects are not included in
\charybdis and, again, our results will be on the conservative side
w.r.t.\ the disappearance of hard jets.

In the final phase the black hole will disappear. Even if a black hole
is produced well above the Planck scale it will evaporate and
eventually enter the Planck region. As previously mentioned, the
\charybdis treatment of this problem is to let the user chose the
number of particles to which the black hole should decay in the end.
For this study we have consistently used 2, partly because this gives
the most continuous transition from a non black hole gravitational
event, and partly since it is the most conservative assumption as it
gives the largest number of very energetic particles.  For the same
reasons we have chosen to define the ``end'' of the evaporation as
whatever happens first of the black hole reaching the Planck mass, or
a decay product having a forbidden momentum as explained in section
\ref{sec:charybdis}.

In total, the ``mistreatment'' by effectively using (Schwarzschild
black hole) Hawking radiation for all phases, terminated by a
two-particle decay, will lead to a maximum radiation of standard model
particles and to a harder spectrum than if additional effects are
taken into account. The simulation is therefore based on conservative
assumptions wrt.\ searching for the disappearance of standard QCD jet
production.

\section{Results}
\label{sec:results}

To investigate possible effects of black-hole production on standard
QCD observables, we have used the \pythia event generator (version
6.227) \cite{Sjostrand:2000wi} for the standard QCD processes together
with the \charybdis program \cite{Harris:2003db} for the production of
black holes and their decays. We have studied predictions for the \Et\ 
spectrum of jets and the dijet invariant mass spectrum at the LHC using a cone
algorithm with a cone radius of 0.7 and a minimum
\Et\ of 250~GeV assuming a calorimeter with $0.1\times0.1$ resolution
in the pseudo rapidity interval $|\eta|<2.5$.

\pythia was set to generate standard QCD events using CTEQ5L parton
distributions \cite{Lai:1999wy}, but for the extra dimensions
scenarios, the differential cross section was cutoff according to
\eqref{eq:qcdcut}.

Black holes were produced according to eq.\ (\ref{eq:sigma}) and
decayed with the \charybdis program, while additional parton showering
was handled by \pythia.\footnote{Hadronization was not included in the
  simulations presented here, but we have checked that our results do
  not change if hadronization is added.}
Apart from the the scale in the parton distribution functions, the varying 
Planck mass, minimal mass, number of extra dimensions and the decision of when
to terminate the decay, the default settings of \charybdis were used.

\FIGURE[t]{%
      \epsfig{file=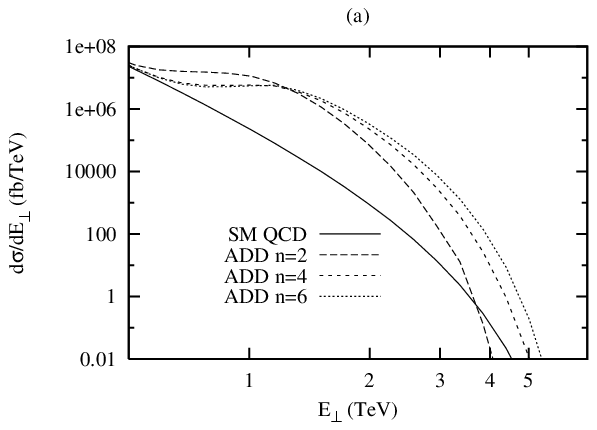,width=7cm}\hfill
      \epsfig{file=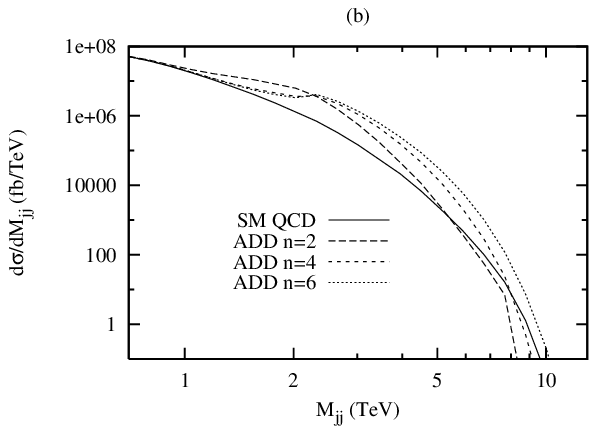,width=7cm}
 \caption{\label{fig:Mp1} Differential jet cross sections for the ADD
   scenario with $\mpl=1$~TeV and $P=1$ for two (long-dashed line),
   four (short-dashed) and six (dotted) extra dimensions, compared
   with the Standard Model prediction (full line). (a) shows the cross
   section as a function of the transverse energy of the hardest jet
   and (b) the cross section as a function of the invariant mass of
   the two hardest jets in an event.}}

In figure \ref{fig:Mp1} the differential jet cross section is shown as
a function of the transverse energy of the hardest jet, and the dijet
cross section as a function of dijet invariant mass for the two
hardest jets, for two, four and six extra dimensions (ADD),
$\mpl=1$~TeV and $P=1$ (corresponding to $\mmin=2.24$~TeV).  Studying
the case of two extra dimensions we find that, although the main
effect is an increase in cross section due to the production of black
holes in the regions just above the Planck mass, the spectra do indeed
fall below the standard model spectra for very large \Et\ and
invariant masses.  The effect should be clearly visible at the LHC
where we would \eg expect over a hundred events with $M_{jj}$ above
8.5~TeV with an integrated luminosity of 100~fb$^{-1}$ (corresponding
to one year of running at $10^{34}~cm^{-2}s^{-1}$) for standard QCD,
while with black hole production there would be basically none. In the
\Et\ spectrum the effect would be harder to see, and only a couple
events expected above 4~TeV from standard QCD would disappear in a
$n=2$ ADD scenario.

However, two extra dimensions is excluded from observations, and if we
increase the number of extra dimensions the temperature rises
according to \eqref{eq:T}. As seen in figure \ref{fig:Mp1} this
results in a drastic increase in the number of hard jets, making it
seem rather unlikely that we will observe the QCD drop at LHC without
further efforts to distinguish QCD and black hole jets. But in
principle it is an enormous effect.  This can be illustrated by
plotting the ratio of QCD events in an ADD world, with \eg 4 extra
dimensions, and QCD events in a four dimensional standard model world.
This is shown in figure \ref{fig:ratio}.

\FIGURE[t]{%
      \epsfig{file=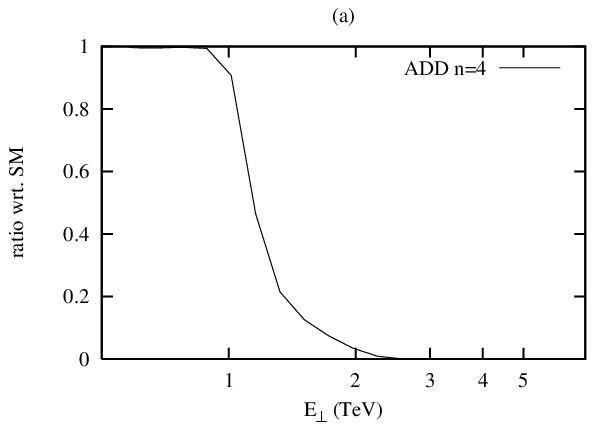,width=7cm}\hfill
      \epsfig{file=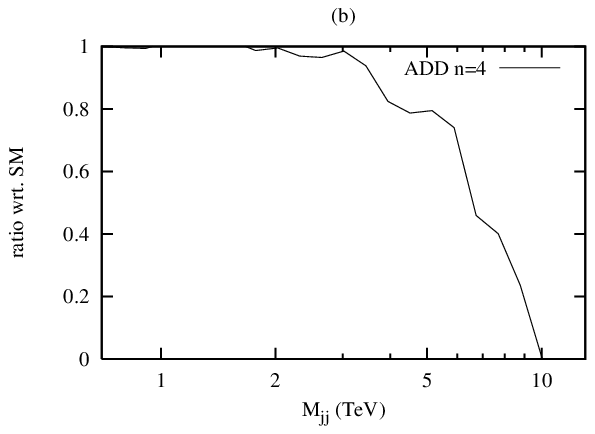,width=7cm}
 \caption{\label{fig:ratio} 
   The number of pure QCD events for 4 extra dimensions, $\mpl=1$
   and $P=1$ divided by the number of QCD events in a Standard Model
   world as a function of (a) the transverse energy of the hardest jet
   and (b) the invariant mass of the two hardest jets in an event.}}

The QCD drop can also be clarified by decomposing the contribution to
the cross section of standard events and black hole decay products as
in figure \ref{fig:components}.  As expected the 'shoulder' in the
spectra is completely dominated by the black hole decay products. The
drop in the QCD cross section is also seen, however, from an
experimental point of view it is completely hidden behind the black
hole decay products.

\FIGURE[t]{%
      \epsfig{file=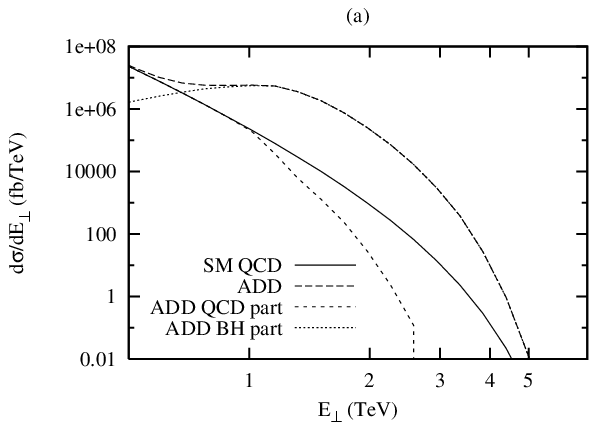,width=7cm}\hfill
      \epsfig{file=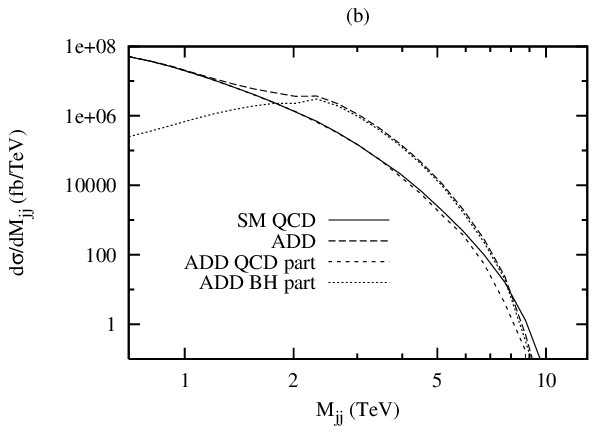,width=7cm}
 \caption{\label{fig:components} 
   Differential jet cross sections for the ADD scenario with
   $\mpl=1$~TeV and $P=1$ for four extra dimensions (long-dashed
   line), compared with the Standard Model prediction (full line).
   Also shown are the contributions from events with (dotted line) and
   without (short-dashed line) black holes. (a) shows the cross
   section as a function of the transverse energy of the hardest jet
   and (b) the cross section as a function of the invariant mass of
   the two hardest jets in an event.}}

Eventually, for high enough energies the QCD drop must appear, since
light black holes cannot decay to particles with energy greater than
half the black hole mass, and heavier black holes typically will
produce less energetic decay products as the temperature is lower.

However, as seen in figure \ref{fig:40TeV} for a imagined 40 TeV
$pp$-collider it turns out that the (large) probability for producing
a heavy black hole, which then emits an unlikely heavy (Boltzmann
suppressed) particle, dominates over the QCD cross section for rather
large transverse energies and invariant dijet masses.

\FIGURE[t]{%
      \epsfig{file=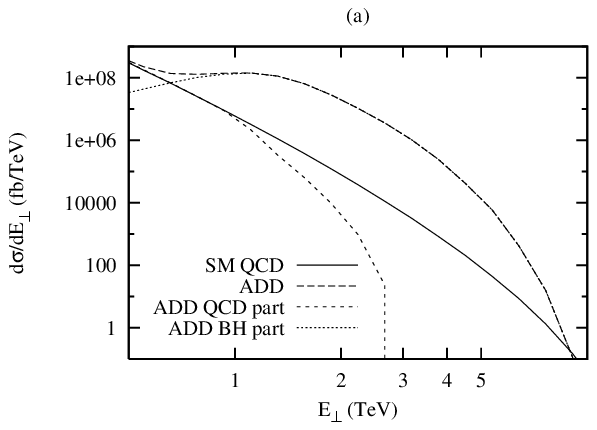,width=7cm}\hfill
      \epsfig{file=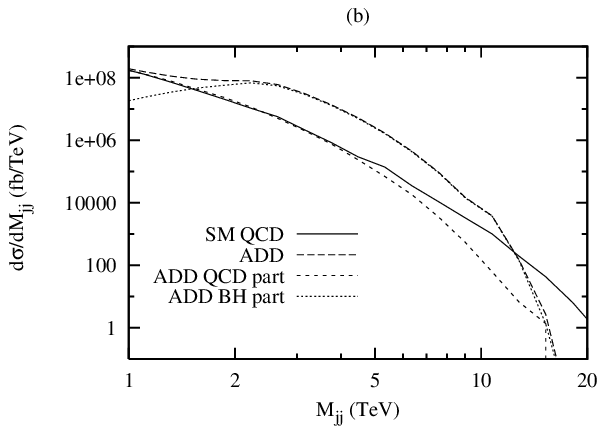,width=7cm}
 \caption{\label{fig:40TeV} 
   Differential jet cross sections at a 40~TeV pp-collider for the ADD
   scenario with $\mpl=1$~TeV and $P=1$ for four extra dimensions
   (long-dashed line), compared with the Standard Model prediction
   (full line).  Also shown are the contributions from events with
   (dotted line) and without (short-dashed line) black holes. (a)
   shows the cross section as a function of the transverse energy of
   the hardest jet and (b) the cross section as a function of the
   invariant mass of the two hardest jets in an event.}}

The effect of varying $P$, and hence also the minimal mass \mmin, is
shown in figure~\ref{fig:Mpdiff} where $P=1$ for $\mpl=1$~TeV (giving
$\mmin=2.24$~TeV) is compared to $P=1/2$ for $\mpl=2$~TeV
($\mmin=2.51$~TeV) and $P=1$ for $\mpl=2$~TeV ($\mmin=4.48$~TeV).
Comparing the lines which have the same $\mpl$ but differ in $P$ and
\mmin\ we see that the extra black holes which are produced if \mmin\ 
is lowered, contribute to the low energy end of the spectrum despite
the fact that they are hotter.  This is because the emitted quantum must 
have an energy less than half the black hole energy. On the other hand, 
the high energy end of the spectra,
where the total cross section eventually would fall below the standard
model QCD cross section, is left more or less unaffected.  This means
that the choice of \mmin\ is not particularly important for determining
the point where the cross section drop would be observed.
Unfortunately this does not imply that it is insensitive to Planck
scale physics as recoil effects on the black hole emitting the
energetic quanta cannot be neglected.

An increased Planck mass results in a later onset of black hole
production and therfore less particles in the low energy end of the
black hole decay spectrum.  On the other hand there are more particles
in the high energy end as the temperature is increased according to
\eqref{eq:T}. For a higher Planck mass it would therfore, as expected,
be more difficult to observe the QCD drop.

\FIGURE[t]{%
      \epsfig{file=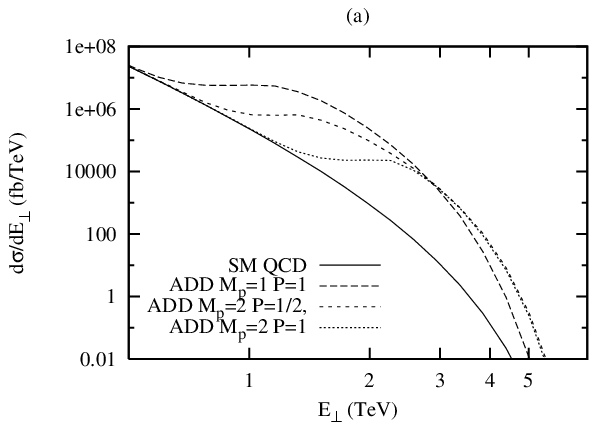,width=7cm}\hfill
      \epsfig{file=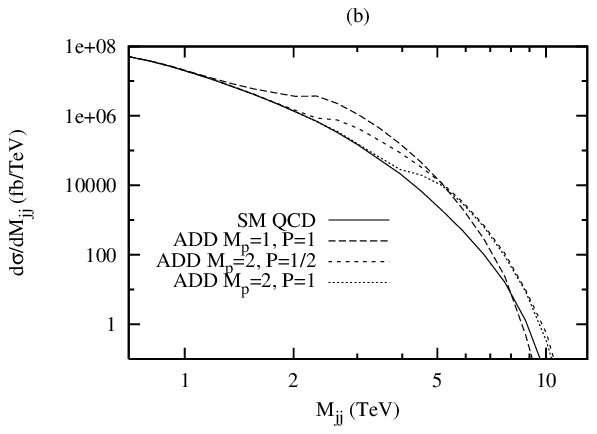,width=7cm}
 \caption{\label{fig:Mpdiff} 
   Differential jet cross sections for the ADD scenario for four extra
   dimensions with $\mpl=1$~TeV and $P=1$ (long-dashed line),
   $\mpl=2$~TeV and $P=1$ (short-dashed) and $\mpl=2$~TeV and $P=1/2$
   (dotted line) compared with the Standard Model prediction (full
   line). (a) shows the cross section as a function of the transverse
   energy of the hardest jet and (b) the cross section as a function
   of the invariant mass of the two hardest jets in an event.}}

One could try to eliminate the extremely energetic black hole events
with a thrust cut to regain the QCD drop, but it does not work
particularly well since the rest of the black hole which emitted the
energetic quanta will have a large momentum in opposite direction. The
event will thus (from a clustering point of view) look like a dijet
event.  Another option is to try a smaller cone radius. This has a
significant effect in the case of 2 extra dimensions, making the drop
clearly visible also in the \Et-spectrum, but turns out to be less
effective in the higher dimensional cases since the black holes there
are hotter.

The overall impression for the ADD scenario is thus that it will be
hard to observe the QCD drop without further efforts to discriminate
between the QCD and black hole radiation. On the other hand we have
made the case worse than it may be in several ways. We have ignored
that some energy will be carried away by invisible gravitational
radiation, thus reducing the observed background from black hole decay
products, and we have maximized the number of energetic particles by
choosing a varying temperature and a 2-body decay in the end of the
evaporation.

If the black holes do not decay on collider timescales, or if a naked 
singularity which decays in the bulk is formed,  
as may be the case in the Randall--Sundrum scenario, there is no radiation 
to camouflage the QCD cross section disappearance. Its disappearance 
may then be a key signal, as shown 
in figure \ref{fig:RS}.    

\FIGURE[t]{%
      \epsfig{file=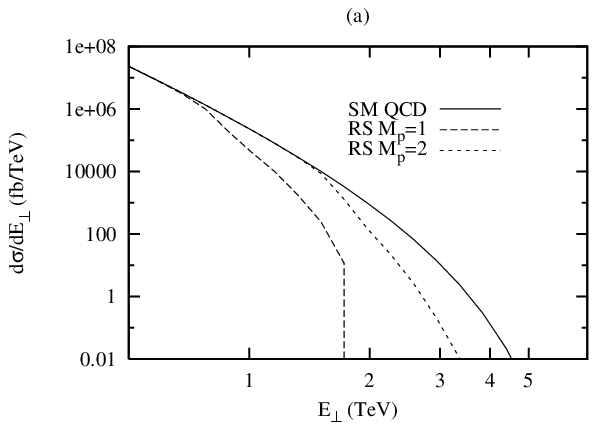,width=7cm}\hfill
      \epsfig{file=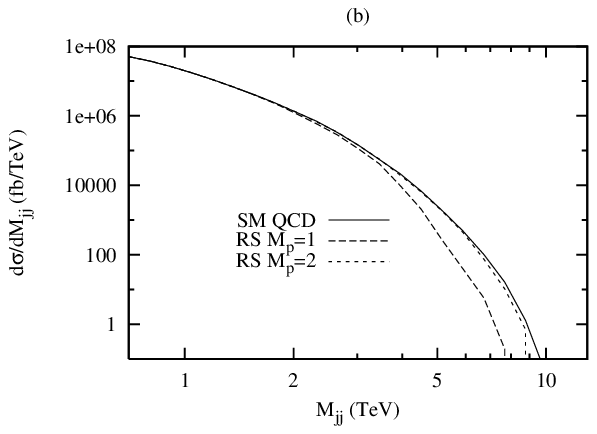,width=7cm}
 \caption{\label{fig:RS}
   Differential jet cross sections for the RS scenario with $P=1$ and
   $\mpl=1$~TeV (long-dashed line) and $\mpl=2$~TeV (short-dashed
   line), compared with the Standard Model prediction (full line). (a)
   shows the cross section as a function of the transverse energy of
   the hardest jet and (b) the cross section as a function of the
   invariant mass of the two hardest jets in an event.}}

%To search for results in the mixture of standard model events and QCD events 
%we use a cone type clustering algorithm to identify the jets. 
%(But the choice of clustering algorithm tunes out to be less important.)
%To simulate the detector environment we also use a rapidity cut of 2.5. 
%The transverse energy cut for the jets to cluster is put to 500?? GeV in 
%the LHC case and to ?? for an imagined 40 TeV pp accelerator. 

\section{Conclusion and outlook}
\label{sec:conclusions}

We have seen that black hole production of partons interacting on a
short enough distance indeed seem to generate a drastic drop in the
QCD cross section at LHC.  However if the created black holes decay on
collider timescales as expected in the ADD scenario, this drop will
(naively) be completely hidden by the black hole decay products even
for rather large transverse energy/invariant dijet mass.  On the other
hand, if the black holes are stable on collider time scales there is
no Hawking radiation to camouflage the QCD drop and the absence of QCD
events may be a key signal.

The point where the extra-dimensional black hole plus standard model
cross section falls below the four dimensional standard model cross
section is sensitive to the number of extremely energetic quanta
emitted by small black holes and this depends on physics in the Planck
region.  While we cannot exclude (due to the large theoretical
uncertainties) that a drop below the standard model QCD-cross section
will be observed at the LHC, it is unlikely (at least in this simple
form) to be important for the identification of the ADD model.
Inventing mores sophisticated methods for distinguishing the black
hole jets from other jets may on the other hand be well worth the
effort.

Finally we point out that it is, in a sense, unphysical to consider
standard model and black hole events without taking gravitational, non
black hole, events into account. Surely the gravitational events will
play a major role close to the Planck mass. For scatterings with
momentum transfer far above the plank mass we expect black hole events
to dominate. We intend to further investigate the effect of
gravitational interaction in future publications.

\bibliographystyle{utcaps}  
%\bibliography{/home/beckett/leif/personal/lib/tex/bib/references,refs} 
\bibliography{references,refs}

\providecommand{\href}[2]{#2}\begingroup\raggedright\begin{thebibliography}{10%
}\itemsep 0mm

\bibitem{Arkani-Hamed:1998rs}
N.~Arkani-Hamed, S.~Dimopoulos, and G.~R. Dvali {\em Phys. Lett.} {\bf B429}
  (1998) 263--272,
\href{http://www.arXiv.org/abs/hep-ph/9803315}{{\tt hep-ph/9803315}}.
%%CITATION = HEP-PH 9803315;%%.

\bibitem{Antoniadis:1998ig}
I.~Antoniadis, N.~Arkani-Hamed, S.~Dimopoulos, and G.~R. Dvali {\em Phys.
  Lett.} {\bf B436} (1998) 257--263,
\href{http://www.arXiv.org/abs/hep-ph/9804398}{{\tt hep-ph/9804398}}.
%%CITATION = HEP-PH 9804398;%%.

\bibitem{Emparan:2000rs}
R.~Emparan, G.~T. Horowitz, and R.~C. Myers {\em Phys. Rev. Lett.} {\bf 85}
  (2000) 499--502,
\href{http://www.arXiv.org/abs/hep-th/0003118}{{\tt hep-th/0003118}}.
%%CITATION = HEP-TH 0003118;%%.

\bibitem{Dimopoulos:2001hw}
S.~Dimopoulos and G.~Landsberg {\em Phys. Rev. Lett.} {\bf 87} (2001) 161602,
\href{http://www.arXiv.org/abs/hep-ph/0106295}{{\tt hep-ph/0106295}}.
%%CITATION = HEP-PH 0106295;%%.

\bibitem{Giddings:2001bu}
S.~B. Giddings and S.~Thomas {\em Phys. Rev.} {\bf D65} (2002) 056010,
\href{http://www.arXiv.org/abs/hep-ph/0106219}{{\tt hep-ph/0106219}}.
%%CITATION = HEP-PH 0106219;%%.

\bibitem{Kanti:2004nr}
P.~Kanti {\em Int. J. Mod. Phys.} {\bf A19} (2004) 4899--4951,
\href{http://www.arXiv.org/abs/hep-ph/0402168}{{\tt hep-ph/0402168}}.
%%CITATION = HEP-PH 0402168;%%.

\bibitem{Banks:1999gd}
T.~Banks and W.~Fischler
\href{http://www.arXiv.org/abs/hep-th/9906038}{{\tt hep-th/9906038}}.
%%CITATION = HEP-TH 9906038;%%.

\bibitem{Han:1998sg}
T.~Han, J.~D. Lykken, and R.-J. Zhang {\em Phys. Rev.} {\bf D59} (1999) 105006,
\href{http://www.arXiv.org/abs/hep-ph/9811350}{{\tt hep-ph/9811350}}.
%%CITATION = HEP-PH 9811350;%%.

\bibitem{Atwood:1999qd}
D.~Atwood, S.~Bar-Shalom, and A.~Soni {\em Phys. Rev.} {\bf D62} (2000) 056008,
\href{http://www.arXiv.org/abs/hep-ph/9911231}{{\tt hep-ph/9911231}}.
%%CITATION = HEP-PH 9911231;%%.

\bibitem{Doncheski:2001se}
M.~A. Doncheski {\em eConf} {\bf C010630} (2001) P314,
\href{http://www.arXiv.org/abs/hep-ph/0111149}{{\tt hep-ph/0111149}}.
%%CITATION = HEP-PH 0111149;%%.

\bibitem{Randall:1999ee}
L.~Randall and R.~Sundrum {\em Phys. Rev. Lett.} {\bf 83} (1999) 3370--3373,
\href{http://www.arXiv.org/abs/hep-ph/9905221}{{\tt hep-ph/9905221}}.
%%CITATION = HEP-PH 9905221;%%.

\bibitem{Myers:1986un}
R.~C. Myers and M.~J. Perry {\em Ann. Phys.} {\bf 172} (1986)
304.
%%CITATION = APNYA,172,304;%%.

\bibitem{Karasik:2003tx}
D.~Karasik, C.~Sahabandu, P.~Suranyi, and L.~C.~R. Wijewardhana {\em Phys.
  Rev.} {\bf D69} (2004) 064022,
\href{http://www.arXiv.org/abs/gr-qc/0309076}{{\tt gr-qc/0309076}}.
%%CITATION = GR-QC 0309076;%%.

\bibitem{Karasik:2004wk}
D.~Karasik, C.~Sahabandu, P.~Suranyi, and L.~C.~R. Wijewardhana {\em Phys.
  Rev.} {\bf D70} (2004) 064007,
\href{http://www.arXiv.org/abs/gr-qc/0404015}{{\tt gr-qc/0404015}}.
%%CITATION = GR-QC 0404015;%%.

\bibitem{Casadio:2001wh}
R.~Casadio and B.~Harms {\em Int. J. Mod. Phys.} {\bf A17} (2002) 4635--4646,
\href{http://www.arXiv.org/abs/hep-th/0110255}{{\tt hep-th/0110255}}.
%%CITATION = HEP-TH 0110255;%%.

\bibitem{Anchordoqui:2002fc}
L.~A. Anchordoqui, H.~Goldberg, and A.~D. Shapere {\em Phys. Rev.} {\bf D66}
  (2002) 024033,
\href{http://www.arXiv.org/abs/hep-ph/0204228}{{\tt hep-ph/0204228}}.
%%CITATION = HEP-PH 0204228;%%.

\bibitem{Sjostrand:2000wi}
{T.~Sjöstrand, and others} {\em Comput.\ Phys.\ Commun.} {\bf 135} (2001)
  238--259,
\href{http://www.arXiv.org/abs/arXiv:hep-ph/0010017}{{\tt
  arXiv:hep-ph/0010017}}.
%%CITATION = HEP-PH 0010017;%%.

\bibitem{Harris:2003db}
C.~M. Harris, P.~Richardson, and B.~R. Webber {\em JHEP} {\bf 08} (2003) 033,
\href{http://www.arXiv.org/abs/hep-ph/0307305}{{\tt hep-ph/0307305}}.
%%CITATION = HEP-PH 0307305;%%.

\bibitem{Harris:2003eg}
C.~M. Harris and P.~Kanti {\em JHEP} {\bf 10} (2003) 014,
\href{http://www.arXiv.org/abs/hep-ph/0309054}{{\tt hep-ph/0309054}}.
%%CITATION = HEP-PH 0309054;%%.

\bibitem{Stump:2003yu}
D.~Stump {\em et al.} {\em JHEP} {\bf 10} (2003) 046,
\href{http://www.arXiv.org/abs/hep-ph/0303013}{{\tt hep-ph/0303013}}.
%%CITATION = HEP-PH 0303013;%%.

\bibitem{Yoshino:2002tx}
H.~Yoshino and Y.~Nambu {\em Phys. Rev.} {\bf D67} (2003) 024009,
\href{http://www.arXiv.org/abs/gr-qc/0209003}{{\tt gr-qc/0209003}}.
%%CITATION = GR-QC 0209003;%%.

\bibitem{Yoshino:2005hi}
H.~Yoshino and V.~S. Rychkov
\href{http://www.arXiv.org/abs/hep-th/0503171}{{\tt hep-th/0503171}}.
%%CITATION = HEP-TH 0503171;%%.

\bibitem{Giddings:2004xy}
S.~B. Giddings and V.~S. Rychkov {\em Phys. Rev.} {\bf D70} (2004) 104026,
\href{http://www.arXiv.org/abs/hep-th/0409131}{{\tt hep-th/0409131}}.
%%CITATION = HEP-TH 0409131;%%.

\bibitem{Rychkov:2004sn}
V.~S. Rychkov
\href{http://www.arXiv.org/abs/hep-th/0410041}{{\tt hep-th/0410041}}.
%%CITATION = HEP-TH 0410041;%%.

\bibitem{Cardoso:2005jq}
V.~Cardoso, E.~Berti, and M.~Cavaglia
\href{http://www.arXiv.org/abs/hep-ph/0505125}{{\tt hep-ph/0505125}}.
%%CITATION = HEP-PH 0505125;%%.

\bibitem{Harris:2004xt}
C.~M. Harris {\em et al.}
\href{http://www.arXiv.org/abs/hep-ph/0411022}{{\tt hep-ph/0411022}}.
%%CITATION = HEP-PH 0411022;%%.

\bibitem{Page:1976ki}
D.~N. Page {\em Phys. Rev.} {\bf D14} (1976)
3260--3273.
%%CITATION = PHRVA,D14,3260;%%.

\bibitem{Frolov:2002xf}
V.~P. Frolov and D.~Stojkovic {\em Phys. Rev.} {\bf D67} (2003) 084004,
\href{http://www.arXiv.org/abs/gr-qc/0211055}{{\tt gr-qc/0211055}}.
%%CITATION = GR-QC 0211055;%%.

\bibitem{Frolov:2003en}
V.~P. Frolov and D.~Stojkovic {\em Phys. Rev.} {\bf D68} (2003) 064011,
\href{http://www.arXiv.org/abs/gr-qc/0301016}{{\tt gr-qc/0301016}}.
%%CITATION = GR-QC 0301016;%%.

\bibitem{Stojkovic:2004hp}
D.~Stojkovic {\em Phys. Rev. Lett.} {\bf 94} (2005) 011603,
\href{http://www.arXiv.org/abs/hep-ph/0409124}{{\tt hep-ph/0409124}}.
%%CITATION = HEP-PH 0409124;%%.

\bibitem{Frolov:2002as}
V.~P. Frolov and D.~Stojkovic {\em Phys. Rev.} {\bf D66} (2002) 084002,
\href{http://www.arXiv.org/abs/hep-th/0206046}{{\tt hep-th/0206046}}.
%%CITATION = HEP-TH 0206046;%%.

\bibitem{Frolov:2002gf}
V.~P. Frolov and D.~Stojkovic {\em Phys. Rev. Lett} {\bf 89} (2002) 151302,
\href{http://www.arXiv.org/abs/hep-th/0208102}{{\tt hep-th/0208102}}.
%%CITATION = HEP-TH 0208102;%%.

\bibitem{Page:1977um}
D.~N. Page {\em Phys. Rev.} {\bf D16} (1977)
2402--2411.
%%CITATION = PHRVA,D16,2402;%%.

\bibitem{Anchordoqui:2002cp}
L.~Anchordoqui and H.~Goldberg {\em Phys. Rev.} {\bf D67} (2003) 064010,
\href{http://www.arXiv.org/abs/hep-ph/0209337}{{\tt hep-ph/0209337}}.
%%CITATION = HEP-PH 0209337;%%.

\bibitem{Lai:1999wy}
{\bf CTEQ} Collaboration, H.~L. Lai {\em et al.} {\em Eur. Phys. J.} {\bf C12}
  (2000) 375--392,
\href{http://arXiv.org/abs/hep-ph/9903282}{{\tt hep-ph/9903282}}.
%%CITATION = HEP-PH 9903282;%%.

\end{thebibliography}\endgroup
\end{document}